\documentclass[cits]{JINST}

\title{The RD53 Collaboration's SystemVerilog-UVM Simulation Framework and its General Applicability to Design of Advanced Pixel Readout Chips}

\author{S. Marconi$^{a,b,c}$, E. Conti$^{b,c}$, P. Placidi$^{b,c}$\thanks{Corresponding author.}~, J. Christiansen$^a$ and T. Hemperek$^d$ \\
\llap{$^a$}CERN,\\
  1211 Geneva, Switzerland\\
\llap{$^b$}Department of Engineering, University of Perugia,\\
  Via G. Duranti 93, I-06125 Perugia, Italy\\
 \llap{$^c$}INFN - section of Perugia,\\
  Via Pascoli, I-06123 Perugia, Italy\\
 \llap{$^d$}Physikalisches Institut, Universit\"at Bonn,\\
  Nu{\ss}allee 12, 531 15 Bonn, Germany\\
  E-mail: \email{pisana.placidi@unipg.it}}

\abstract{The foreseen Phase 2 pixel upgrades at the LHC have very challenging requirements for the design of hybrid pixel readout chips. A versatile pixel simulation platform is as an essential development tool for the design, verification and optimization of both the system architecture and the pixel chip building blocks (Intellectual Properties, IPs). This work is focused on the implemented simulation and verification environment named VEPIX53, built using the SystemVerilog language and the Universal Verification Methodology (UVM) class library in the framework of the RD53 Collaboration. The environment supports pixel chips at different levels of description: its reusable components feature the generation of different classes of parameterized input hits to the pixel matrix, monitoring of pixel chip inputs and outputs, conformity checks between predicted and actual outputs and collection of statistics on system performance. The environment has been tested performing a study of shared architectures of the trigger latency buffering section of pixel chips. A fully shared architecture and a distributed one have been described at behavioral level and simulated; the resulting memory occupancy statistics and hit loss rates have subsequently been compared.}


\begin{document}

\section{Introduction}

The development of next generation pixel readout Integrated Circuits (ICs) will be required to enable the ATLAS and CMS Phase 2 pixel upgrades in the High Luminosity - Large Hadron Collider (HL-LHC) at CERN \cite{bibloi1, bibloi2}. The design of such ICs presents major challenges including improved resolution, much higher hit rates ($1-2\,\mbox{GHz/cm}\textsuperscript{2}$), unprecedented radiation level ($10\,\mbox{MGy}$), much higher output bandwidth and low power consumption. The RD53 Collaboration was established to meet these challenges \cite{bib1}. In addition, it will be required to include more logic and more buffering, increasing the complexity of the system, that will be achieved thanks to the scaling of the technology node foreseen for third generation pixel readout chips \cite{bib2}. 
System design will be more and more complicated due to the strict requirements and its functional verification will be a crucial step as 
it will be much more difficult to catch all possible functional errors.

In the challenging context of the ATLAS and CMS Phase 2 pixel upgrades, a flexible and generic pixel simulation and verification platform is as an essential development tool. IC designers from different experiments and institutes, building different Intellectual Properties (IPs) implementing different functionalities, can benefit from a single simulation framework for effectively performing complex optimization of the system architecture and verifying multiple IPs integration and on-chip inter-IP communication. It could also be a valuable design tool for other pixel applications, e.g. for the vertex detectors of the Linear Collider Detector (LCD) project \cite{biblcd}.
Moreover, the framework will also be used at further stages of verification and testing of the RD53 designed chips.

System-level design and verification are currently done in industry through complex environments based on common verification methodologies like Open Verification Methodology (OVM) and Universal Verification Methodology (UVM) \cite{bib3}, built on top of the the hardware description and verification language 
SystemVerilog \cite{bib4}. On one hand, SystemVerilog supports multiple chip description levels and this is vital for simulating and verifying systems from very high level to very detailed gate-level; on the other hand, 
the UVM library provides a set of documented base classes for all the building blocks of the environment and highly customizable reporting features. Therefore, one can profit from SystemVerilog and UVM for the development of reusable verification environments to be used by multiple designers with different needs and goals. Such verification methodologies are becoming a consolidated standard \cite{bib4a, bib4b, bib4c} and the High Energy Physics community has also started to adopt them. To provide some examples, 
such methodologies have been used for the design and verification of Velopix and Timepix3 \cite{bib5} while the ATLAS pixel chip FE-I4 has been verified through a dedicated OVM environment \cite{bib6}.

This paper presents a new and efficient Verification Environment for PIXel chips developed in the framework of the RD53 Collaboration (VEPIX53) \cite{bib1} facing the challenging requirements related to the Phase 2 upgrades. Such a simulation environment, based on SystemVerilog and UVM, performs the test of the architecture and IPs by reusing test bench and test scenarios. This paper is organized as follows: Section 2 describes the structure of the framework and its components; Section \ref{sec:hitgen} focuses on the stimuli generator, which provides a set of parameterized detector hits for the pixel chip simulation, and on the currently implemented test scenarios. An application of the environment related to a particular test case, such as the study of pixel chip buffering architectures at behavioral level, is described in Section \ref{sec:testcase}, while Section \ref{sec:conclusions} draws the conclusions and presents further developments.

\section{The VEPIX53 simulation and verification framework}

The goal of the RD53 simulation working group is to develop a flexible pixel verification environment with the main requirement of simulating alternative pixel chip architectures at increasingly refined level as the design progresses. In order to perform global architecture evaluations and all the incremental extensions and refinements of a final design, large sets of pixel hits and triggers need to be produced in a fully automated fashion. In particular, it should be possible to do the following:
\begin{itemize}
\item generate large data sets of pixel hits and triggers with given constrained random distributions within the framework;
\item import particle hits from external full detector/experiment Monte Carlo simulations and detailed sensor simulations and mix them with statistically generated hits;
\item specify directed tests for debugging specific functions and problems and verifying extreme cases.
\end{itemize}
In order for the environment to be used among the collaboration, it must also be able to support IP development: designers should be able to plug their IP to the environment with minimal effort and verify it at multiple levels of description.

\begin{figure}[htbp]
  \centering
  \includegraphics [width=\textwidth]{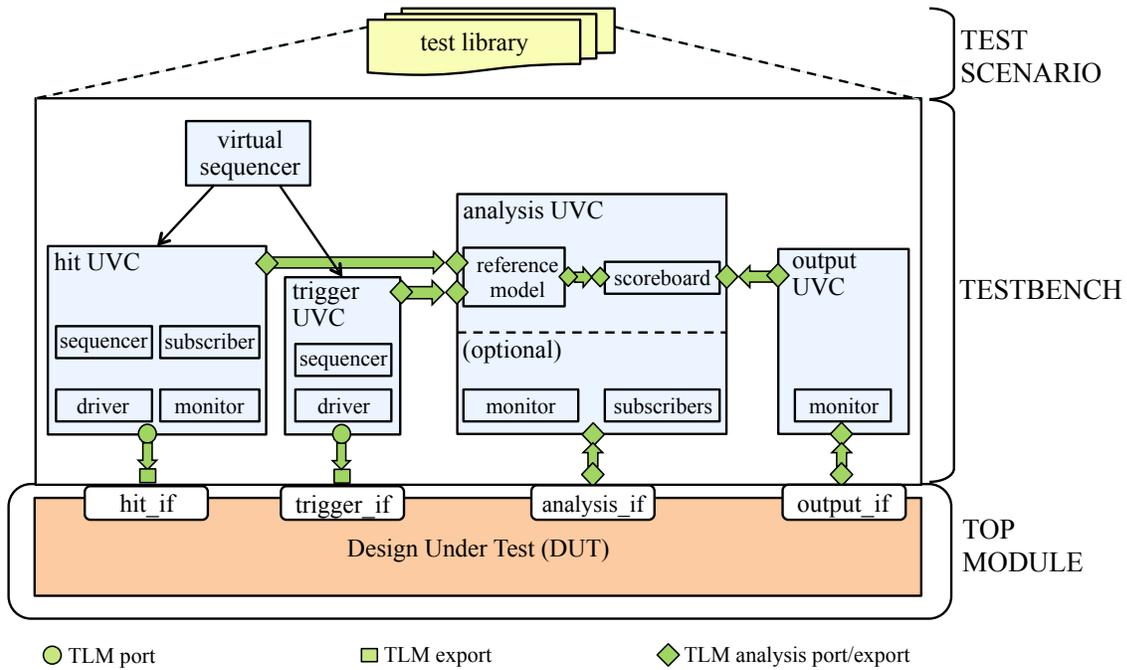}
  \caption{Block diagram of the VEPIX53 simulation and verification environment.}
  \label{fig:topdiagram}
\end{figure}

An initial version of VEPIX53 has been made available to RD53 community. 
As reported on the block diagram in Figure~\ref{fig:topdiagram}, the environment is divided in three main parts: %
 \emph{i)} a top module, which contains the Design Under Test (DUT) and hooks it with the rest of the environment through interfaces in order to build a layered test structure; %
 \emph{ii)} a testbench, which includes all the UVM Verification Components (UVCs) and the \emph{virtual sequencer}; %
 \emph{iii)} a test scenario portion which defines the configuration of the UVCs and describes the tests that are performed during simulations by specifying the constraints to the input stimuli to be sent to the DUT. %
While it is possible to have different levels of description for the DUT, the testbench is implemented at the Transaction Level Modeling (TLM) level 
and the UVCs together with the components in the test scenario are inherited from the UVM class library. TLM components feature a higher level of abstraction 
where the communication between such components takes place through high-level channels 
with transaction objects.
The boundary shown in Figure~\ref{fig:topdiagram} between the DUT and the top-level environment is generic as the environment describes the interfaces at a different level, depending on the DUT. 

\subsection{Top module, interfaces and transactions}
\begin{figure}[htbp]
  \centering
  \includegraphics [width=\textwidth]{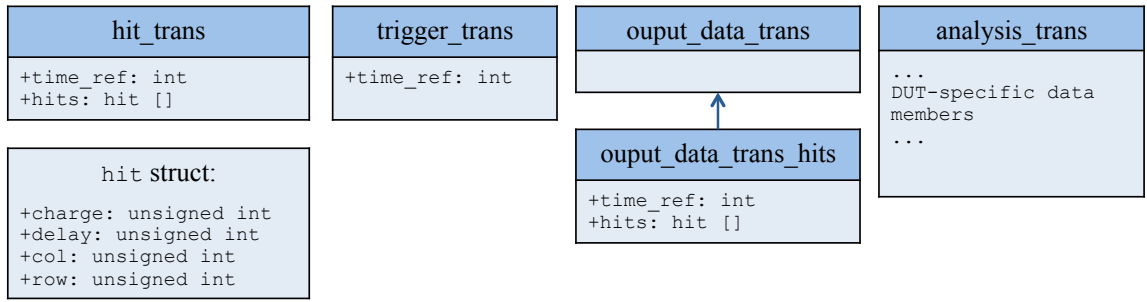}
  \caption{Unified Modeling Language diagrams of the transaction classes defined for VEPIX53.}
  \label{fig:transactions}
\end{figure}

Based on previous generation pixel chips (e.g. \cite{bib5}), common interfaces and corresponding transactions for communication between DUT and testbench have been identified for VEPIX53 (Figure~\ref{fig:transactions}): hit, trigger, output and analysis.

The hit interface (\emph{hit\_if}) includes the signal generated in the pixel sensor matrix due to particles crossing the detector. Hit transactions that transit in the corresponding TLM channel are composed of a time reference field (that refers to the bunch crossing cycle in the HL-LHC) and of an array of incoming hits described by the following fields: charge, delay (with respect to the bunch crossing cycle) and identification of the pixel in the matrix, given through column and row addresses.
 
The second interface: \emph{trigger\_if}, is provided in the case the DUT is a pixel chip that performs selection of events of interest by using an external trigger signal. The corresponding trigger transaction contains just a time reference.

A separate interface named \emph{output\_data\_if} is available for the DUT output. Different subcategories of output transaction can be identified (e.g. processed hits or register data) and the currently defined one, focused on the core task of the pixel chip, contains the same data members of the incoming hit transaction.
 
For monitoring the internal status of the DUT and collecting statistics on chip performance we have defined an optional interface, named \emph{analysis\_if}, containing internal DUT information to be monitored. The corresponding analysis transaction includes 
internal information and is therefore DUT-specific.

All interfaces also contain signals used for synchronization of the DUT and of the environment itself, i.e. mainly the clock. The standard and modular approach that has been followed in the design of the UVCs that interact with each interface does not prevent the addition of new interfaces. For instance, it is planned to enhance the framework with a configuration and a testing interface.

It should be underlined that with the intent of implementing a tool to be used at different steps of the design progress, not only interfaces have been described at TLM level, but also an optional conversion into physical signals and vice versa is provided.

\subsection{UVM Verification Components}
In the testbench, as shown in Figure~\ref{fig:topdiagram}, we defined four different UVCs, each devoted to the respective interface, and a \emph{virtual sequencer}.

The hit and trigger UVCs include objects inherited from the \emph{uvm\_sequencer} and \emph{uvm\_driver} classes. These components provide constrained random stimuli to the DUT (pixel chip matrix and trigger logic) and are driven by the \emph{virtual sequencer}, which coordinates the stimulus across different interfaces and the interactions between them (the adjective \emph{virtual} is related to the fact that the component is not directly linked to a DUT interface). A more detailed description of the hit generation will be provided in Section \ref{sec:hitgen}.

The output UVC includes an entity, inherited from the \emph{uvm\_monitor} class, which reads the DUT outputs through the \emph{output\_data\_if} interface and packs them in the output data hit transaction.

The analysis UVC collects transactions from all the other components. Two main objects, inherited from the UVM class library, are therein defined: a reference model and a scoreboard. The former predicts the expected DUT output from hit and trigger transactions, while the latter compares the predicted and actual DUT outputs providing error/warning/information messages, depending on the verification process result, together with a final summary on total observed matches and mismatches. The analysis UVC also builds transactions from the optional analysis interface using a dedicated object inherited from the \emph{uvm\_monitor} class. These transactions are used not only as a support to the prediction of the DUT output in the reference model, but also for the collection of statistics on the DUT performance. This is done by additional objects inherited from the \emph{uvm\_subscriber} class. Further details are described for the particular test case in Section \ref{sec:testcase}.

\section{Stimuli generator for pixel hits emulation}
\label{sec:hitgen}

The stimulus to the DUT is generated by creating dedicated sequences, inherited from the \emph{uvm\_se\-quence} class, named \emph{hit\_cluster\_sequence}. In the tests defined in the test scenario the main field of the sequence is set, profiting from the UVM configuration database facility, and the \emph{virtual sequencer} launches the generation of consistent transactions in the sequencers contained in the hit and trigger UVCs. Such transactions are passed to the drivers, translated and subsequently injected to the DUT.
\subsection{Configuration of different classes of hits}
\begin{figure}[htbp]
  \centering
  \includegraphics [width=\textwidth]{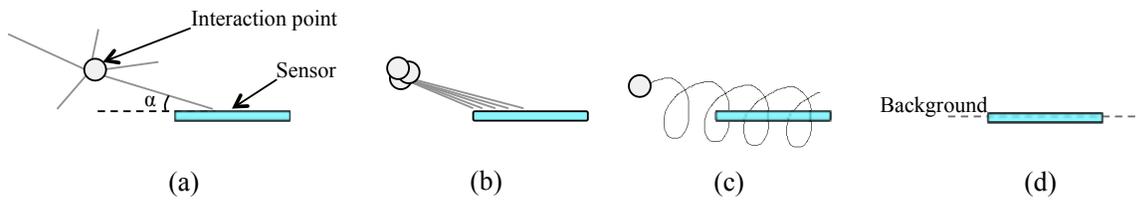}
  \caption{Classes of hits generated by VEPIX53: (a) single charged particles, (b) jets, (c) loopers, (d) machine background particles.}
  \label{fig:hitclasses}
\end{figure}

\begin{figure}[htbp]
  \centering
  \includegraphics [width=\textwidth]{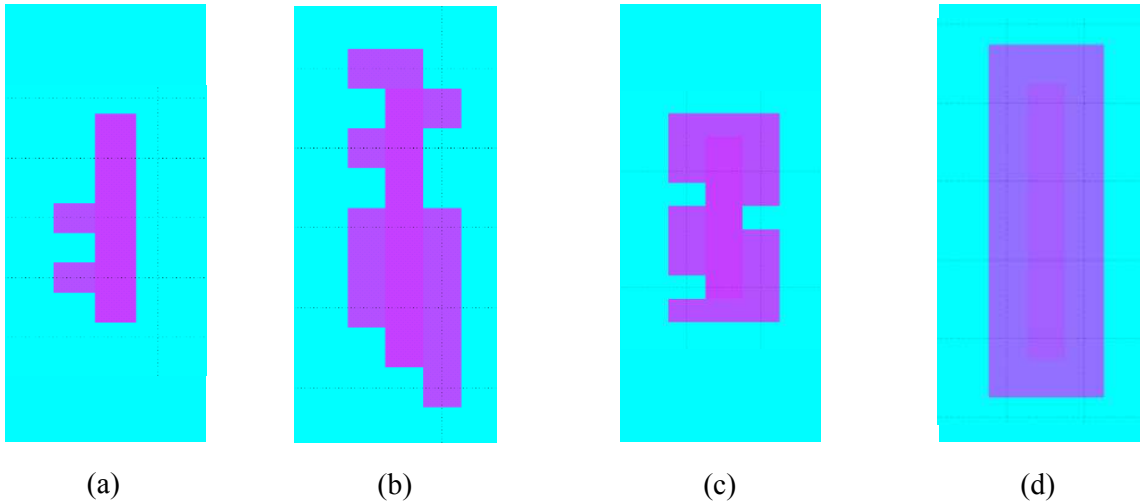}
  \caption{Examples of elongated clusters with charge sharing and cross-talk. The percentages of hit pixels (surrounding the central cluster) are: (a) 10\%, (b) 50\%, (c) 80\%, (d) 100\%.}
  \label{fig:track_CS2}
\end{figure}

\begin{figure}[htbp]
  \centering
  \includegraphics [scale=1]{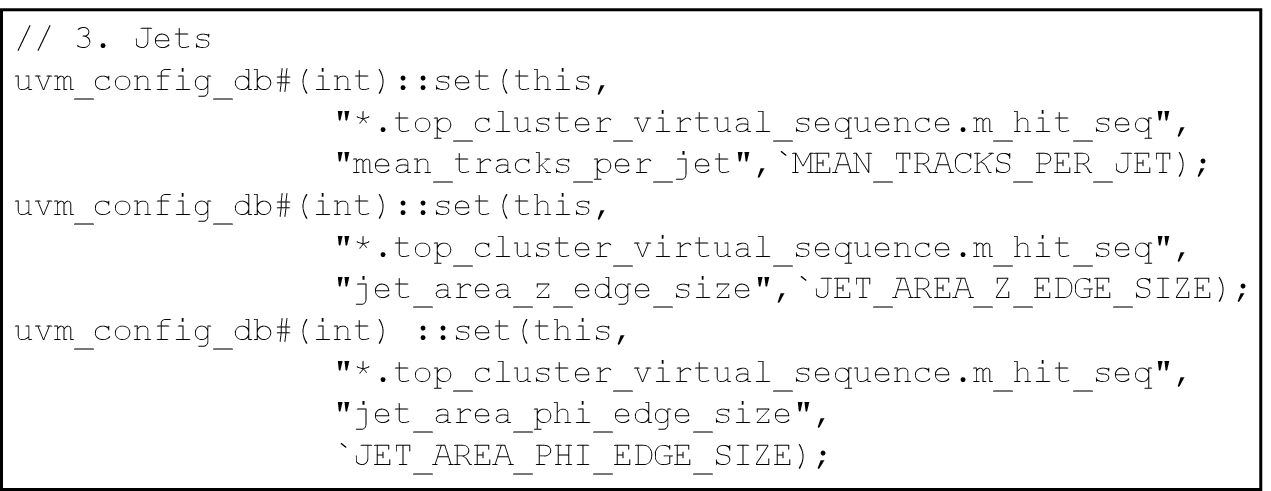}
  \caption{Test code used to configure the sequence parameters related to the jets through the UVM configuration database (similar code can be found for all the other sensor and hit parameters that have been listed).}
  \label{fig:code_snippet}
\end{figure}

The detector hits have been classified on the base of expected pixel hits at the HL-LHC. The following classes of hits (represented in Figure~\ref{fig:hitclasses}) have been identified: 
\begin{enumerate}
\item charged particles (Figure~\ref{fig:hitclasses} (a)) crossing the sensor at a given angle. Each particle fires a variable number of pixels, generating a cluster of pixel hits, the average size of which depends on the angle formed by the particle with respect to the sensor;
\item jets (Figure~\ref{fig:hitclasses} (b)), collimated bunches of final state hadrons coming from hard interactions \cite{bib10}. Such a phenomenon has been modeled as a combination of multiple and close particles that hit the sensor, giving rise to a group of clusters localized in a certain area of the matrix;
\item loopers (Figure~\ref{fig:hitclasses} (c)), associated to soft (low energy) charged particles that in the solenoidal magnetic field become curling tracks \cite{bib10}. This class of hits has not yet been investigated in detail, so in the current version of the environment they have been modeled as single particles crossing the sensor with an angle of $90\,^{\circ}\mathrm{}$ since they are likely to hit the sensor perpendicularly to its surface;
\item  machine background particles (Figure~\ref{fig:hitclasses} (d)), that can be seen as particles very close to be parallel to the sensor surface, generating very large elongated clusters;
\item noise hits, from electronics noise in the analog pixel front-end and soft X rays.
\end{enumerate}
Such interactions are modeled by taking into account the shape of the cluster of fired pixels. Several parameters can be set in order to efficiently drive the generation of the different classes of hits: 
\begin{itemize}
\item generic sensor parameters (e.g. pixel pitch and sensor thickness), common to all the classes of hits;
\item the range of possible amplitudes for the charge deposited by each hit (common to all the classes of hits);
\item a rate for each class of hit, specified in Hz/cm\textsuperscript{2}. A corresponding number of single charged particles, loopers, jets, machine background particles and noise hits is generated accordingly at each bunch crossing cycle;
\item specific parameters for single charged particles are used for modeling the cluster shape left by them. A particle angle parameter is used to compute the number of pixels in the core of the cluster and it is related to the position of the sensor in the whole pixel detector; a second parameter defines the amount of fired pixels surrounding the core of the cluster, in order to take into account in a very simplified way the combined effect of charge sharing, cross-talk and charge deposition between pixels. The hit generation algorithm performs a constrained randomization of the amplitude value and assigns it to the pixels in the core of the cluster. If the second parameter set is not zero, additional surrounding pixels are also hit (with slightly lower amplitude) and their percentage can be defined. The shape of the generated cluster is in reality also affected by several quantities and conditions that are not reproduced in the SystemVerilog framework (e.g. magnetic field, partial depletion operation, possible higher operation voltages to correct radiation damage effects, and delta rays
) and therefore the defined parameters are intended to provide realistic-looking clusters rather than a physical model. In Figure~\ref{fig:track_CS2} clusters generated with a fixed particle angle but variable cross-talk are shown;
\item specific parameters for jets are the average number of charged particles forming a jet and the size of the area in which they are concentrated (the related portion of code is reported in Figure~\ref{fig:code_snippet} as a example of how parameters are settable through the test);
\item a specific parameter for the monster class of hits is the direction along which the pixels of the matrix are fired (either hitting a full column or a full row).
\end{itemize}

%
It is planned to enhance the stimuli generation in order to not only statistically generate hits, but also import particle hits from external full detector/experiment Monte Carlo simulations and detailed sensor simulations and mix them. Directly interfacing the hit generator with the data analysis framework ROOT is also being considered \cite{bib10b} as future development, since this tool is extensively used in the HEP community.

\subsection{Monitoring and statistics collection}
The hit UVC is also in charge of monitoring the actually produced stimuli for multiple purposes: debug, graphical representation as well as statistics on the generated data. To this end, a dedicated component inherited from the \emph{uvm\_subscriber} class has been instantiated in the UVC. High flexibility is provided to the user by configuring the behavior of the subscriber. In detail it can be chosen:
\begin{itemize}
\item to write all generated hits at each bunch crossing cycle to a \emph{.dat} file, which contains information on the position in the matrix and the amplitude;
\item to build histograms on generated hits. These can be either printed every given number of bunch crossing cycles into a \emph{.dat} output file (for further elaboration or graphical visualization) or used in the framework itself to assess the actual hit rate all over the simulation (and finally print it to a text file).
\end{itemize}
Moreover, it is possible to monitor hit rate at different levels of details by setting a configuration object according to the following options:
\begin{itemize}
\item NONE, which means that the dedicated \emph{statistics.txt} file is not generated at all;
\item ONLY\_TOTAL, that causes the file to contain only information on observed total hit rate per cm\textsuperscript{2} (possibly originating from different sources);
\item DETAILED, for obtaining both information on total hit rate per cm\textsuperscript{2} and classified hit rate per cm\textsuperscript{2}, depending on the source (i.e. charged particles, loopers, jets, monsters, noise hits). 
\end{itemize}
Also for hit monitoring, a future development being considered is the implementation of a link to ROOT that would provide extensive and online statistical analysis.

\section{Test case: study of buffering architectures}
\label{sec:testcase}
The VEPIX53 verification environment has been used for studying the local storage section of pixel chips. Trigger latency buffers are critical blocks for the Phase 2 upgrade, as they are going to be used for handling the high data rate resulting from a planned pixel hit rate of $1-2\,\mbox{GHz/cm}\textsuperscript{2}$ and longer trigger latency. Therefore it is necessary to optimize such a storage in order to achieve compact circuitry and low power consumption. A solution that has been adopted for pixel chips, such as FE-I4 for the ATLAS experiment or Timepix3, consists of grouping the pixel unit cells (PUCs) in regions (pixel regions, PRs), where the data buffering is shared.

\begin{figure}[htbp]
  \centering
  \includegraphics [width=.85\textwidth]{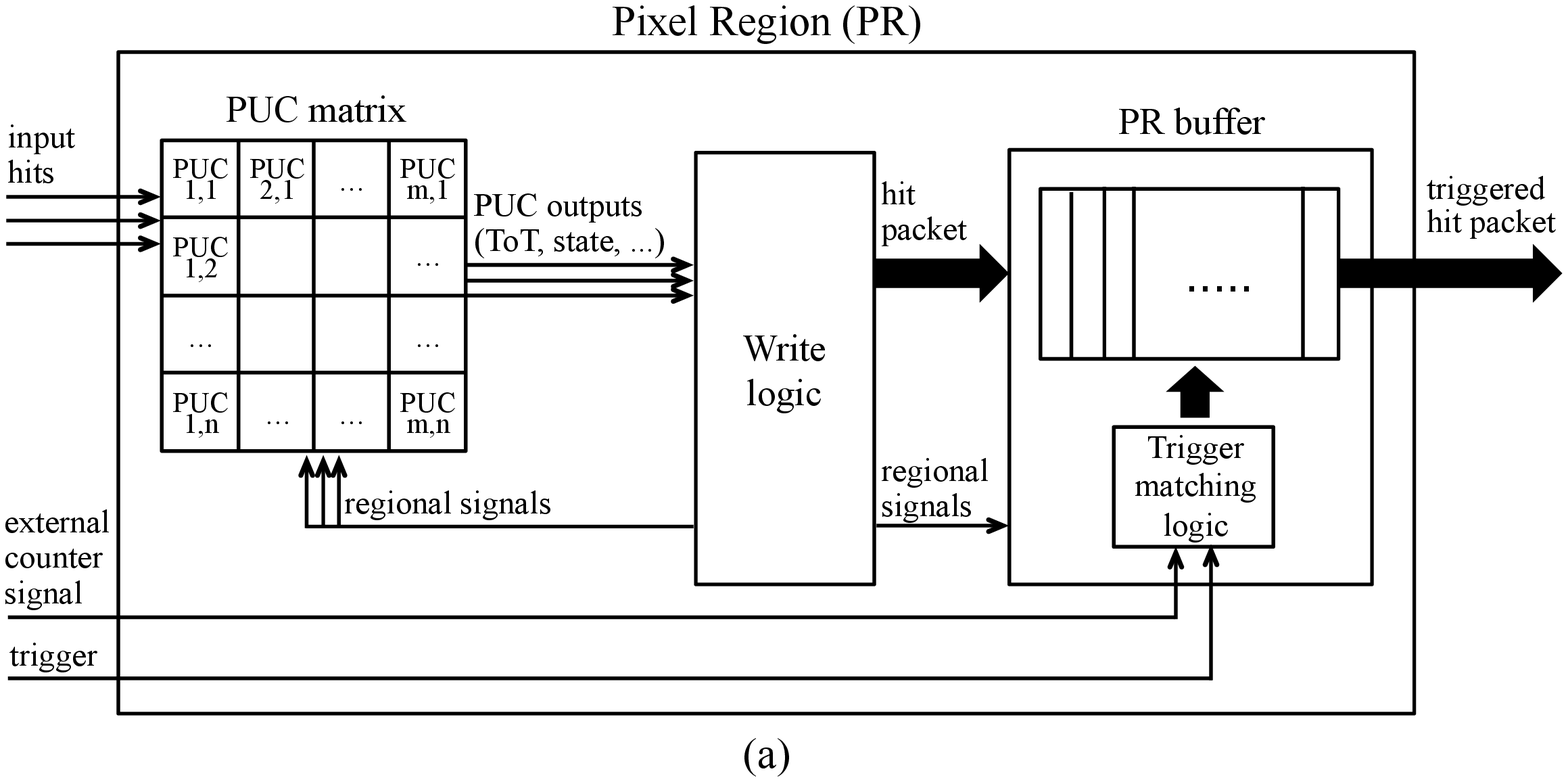}
  \includegraphics [width=.85\textwidth]{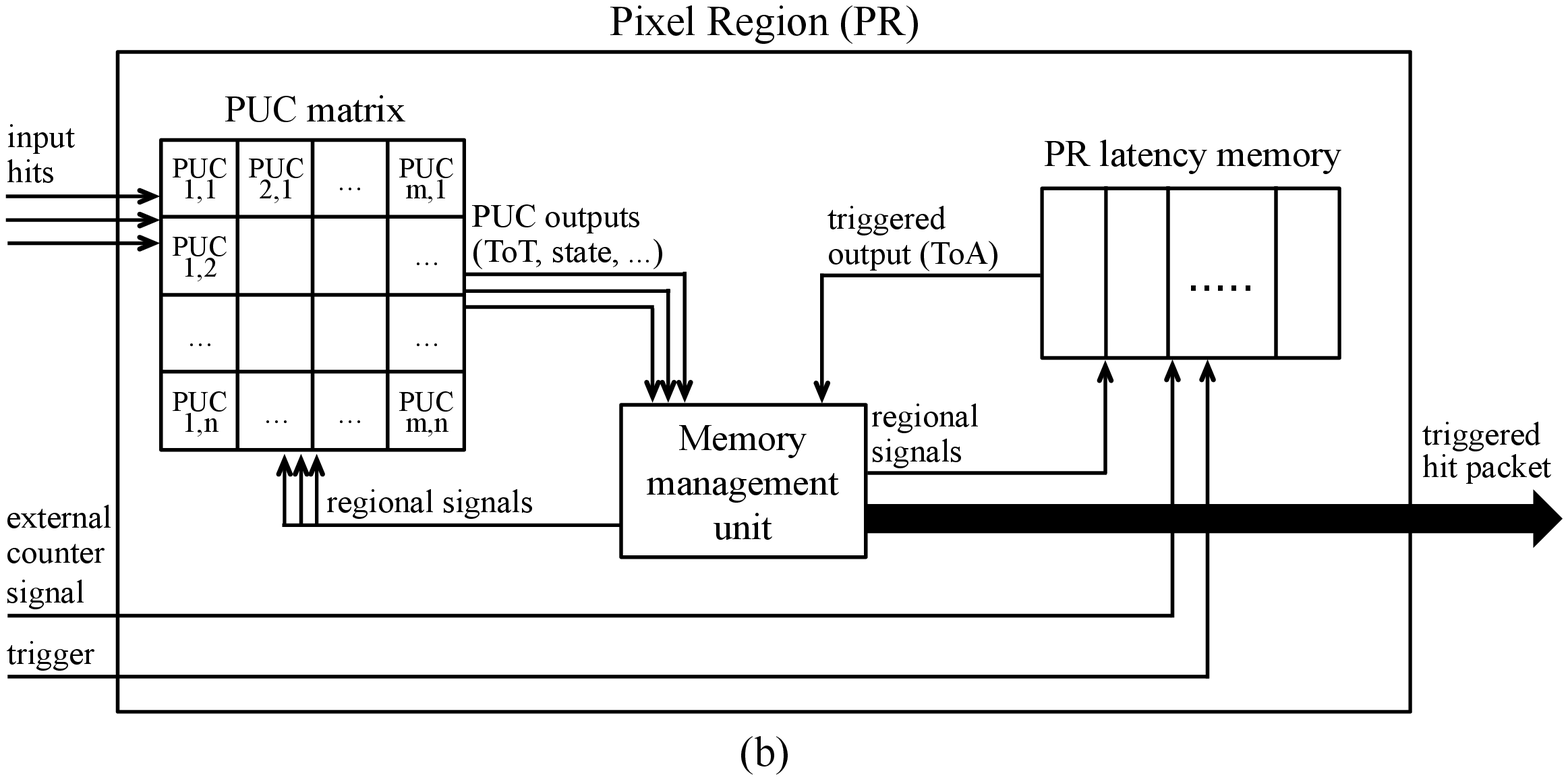}
  \caption{(a) Block diagram of the zero-suppressed FIFO buffering architecture; (b) block diagram of the distributed latency counters buffering architecture. It should be underlined that in this architecture there is a ToT buffer in each PUC.}
  \label{fig:bufferingarchitectures}
\end{figure}

For the study described in this Section we have investigated the performance of two different buffering architectures for PRs: %
\begin{itemize}
\item zero-suppressed FIFO, where the hit information (time and charge) is stored in a unique shared buffer;
\item distributed latency counters (adopted in the FE-I4 \cite{bib10a}), where independent PUC buffers store hit charge and a shared latency memory stores hit time.
\end{itemize}
The architectures have been described at behavioral level (block diagrams are reported in Figure~\ref{fig:bufferingarchitectures}) as part of a DUT which contains a PR of parameterized size. %
Each PUC of the region features a charge converter module, which abstracts the behavior of the analog front-end by converting the hit charge into a discriminator output, and a counter which measures the hit Time over Threshold (ToT). The hit Time of Arrival (ToA), on the other hand, is provided to the PR as a signal coming from a counter module defined at the end of column sector of the pixel chip.
 
In the zero-suppressed FIFO architecture (Figure~\ref{fig:bufferingarchitectures} (a)) the regional buffer stores hit packets made of a shared ToA value, followed by the ToT values from each PUC; trigger matching is checked by comparing the external counter signal, subtracted by the trigger latency, with the stored ToA.

In the distributed latency counters architecture (Figure~\ref{fig:bufferingarchitectures} (b)), instead, the arrival of the hit enables latency down counters, defined inside each cell of the latency memory, and trigger matching is checked when such counters reach zero (i.e. after the latency); a memory management unit links the read and write pointers to the memory cells among the PUC buffers and the latency memory and assembles the triggered hit packets with the same arrangement as that of the previously described architecture. In both the architectures a $40\,\mbox{MHz}$ bunch crossing clock is provided to the PR.

\begin{figure}[htbp]
  \centering
  \includegraphics [width=.85\textwidth]{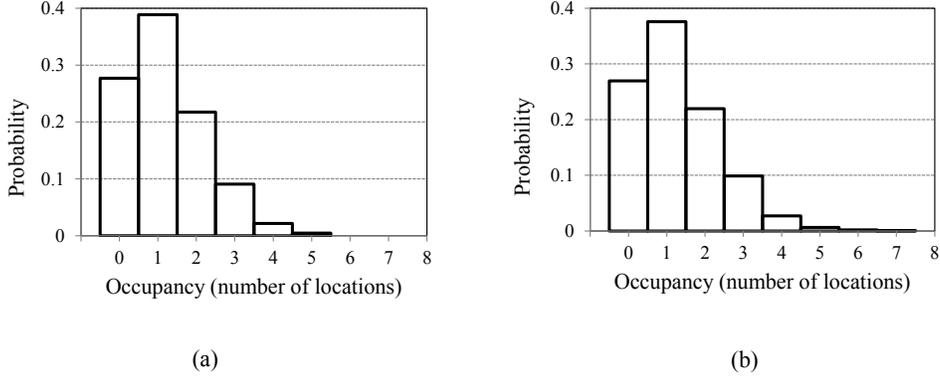}
  \caption{Examples of buffer occupancy histograms carried out from the output file generated by the buffer subscriber for a PR of size 2$\times$2 pixels featuring (a) a zero-suppressed FIFO buffering architecture and (b) a distributed latency counters architecture.}
  \label{fig:occupancyhistograms}
\end{figure}

In order to assess the buffering performance we used the analysis UVC of VEPIX53 to collect statistical information on memory occupancy and lost hits: dedicated internal signals have been defined in the DUT as part of the \emph{analysis\_if} interface. The corresponding analysis transactions that are built in the UVC are forwarded to two objects inherited from the \emph{uvm\_subscriber} class: a buffer subscriber, which produces at the end of the simulation an output file containing the histogram of the PR buffer occupancy (an example is reported in Figure~\ref{fig:occupancyhistograms}, and a lost hits subscriber, which produces an output file summarizing the amount of lost hits in the PR with additional information on the source of hit loss. 

\subsection{Simulation results}
\begin{figure}[htbp]
  \centering
  \includegraphics [width=.5\textwidth]{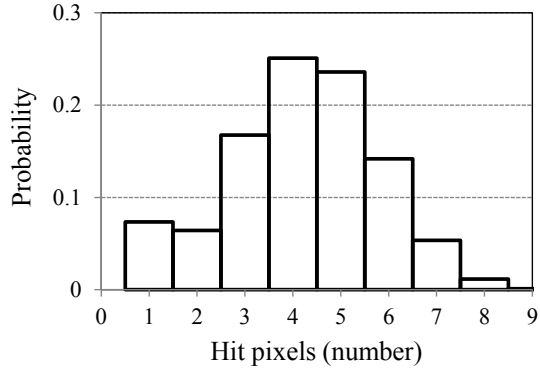}
  \caption{Distribution of the number of hit pixels per cluster generated during the simulation (average:~4.215~hit~pixels).}
  \label{fig:simulationdistribution}
\end{figure}

The two architectures have been simulated using single charged particles as input hits for the DUT, crossing with an angle of $90\,^{\circ}\mathrm{}$ (i.e. particles perpendicular to the detector) and 43\% fired pixels surrounding the core of the cluster: this corresponds to the typical distribution of the number of hit pixels per cluster depicted in Figure~\ref{fig:simulationdistribution}. A pixel size of $50\times50\,\mu\mbox{m}\textsuperscript{2}$ has been used for the simulation. The particle rate has been set in order to obtain a total hit rate of $2\,\mbox{GHz/cm}\textsuperscript{2}$, while the trigger latency has been fixed to $10\,\mu \mbox{s}$, compatible with the expected ones for the Phase 2 upgrade. Furthermore, a uniform distribution has been set for the hit charge which is such that the resulting processed ToTs are uniformly distributed in the full range of amplitude [0:16], corresponding to [0:$400\,\mbox{ns}$]. 
%
%
It should be underlined that because the ToT amplitude is randomized, there is a non-negligible probability of obtaining low amplitude values for the pixels in the core of the cluster: in this case no pixels are fired in the surrounding area. This justifies the probability value associated to ToT equal to 1 in the distribution in Figure~\ref{fig:simulationdistribution}, which would have been significantly lower if clusters were generated without taking the amplitude value into account. %
The simulations have been run for 500,000 bunch crossing cycles for square PRs with size from 1$\times$1 to 8$\times$8 pixels. All the buffers in both the architectures have been oversized in order to collect statistics on the occupancy.

\begin{figure}[htbp]
  \centering
  \includegraphics [width=\textwidth]{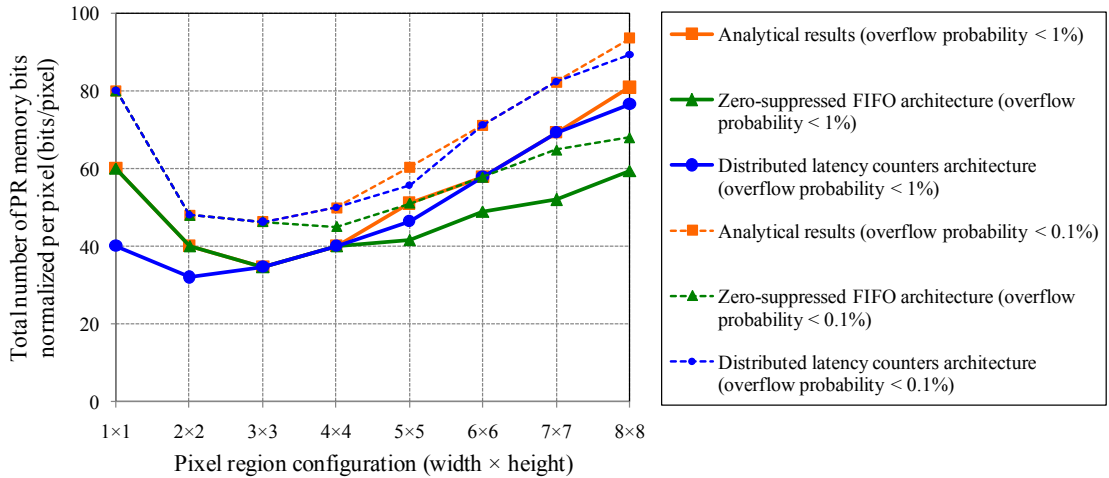}
  \caption{Total required number of memory bits per pixel as a function of PR size for the zero-suppressed FIFO (in green) and distributed latency counters (in blue) buffering architectures, compared with previously presented analytical results \cite{bib11} (in orange), with respect to a buffer overflow probability of 1\% (solid line) and 0.1\% (dashed line).}
  \label{fig:memorybits}
\end{figure}

From the histograms produced by the buffer subscriber we have carried out
the required number of locations for keeping the buffer overflow probability below 1\% and 0.1\%, which are acceptable values for the high rate regions of the pixel detector. This quantity, multiplied by the number of bits of a single location, gives the total required number of memory bits in a PR and is then normalized with respect to the number of pixels in the PR. The results have been calculated for each PR configuration with the two different architectures and have also been compared with those obtained analytically in a previous study using the same input conditions \cite{bib11}. By assuming a hit packet where the ToA is represented using 16 bits and 4 bits are used for each ToT, we have obtained the plots represented in Figure~\ref{fig:memorybits}. These plots show that the two buffering architectures have correctly been implemented at behavioral level and the required number of memory bits is comparable between them. Furthermore, both the simulation results and the analytical ones are confirming that square PRs with size from 2$\times$2 to 4$\times$4 pixels are the most promising. The choice of the most appropriate configuration during design will depend on the storage cells that will be used in the designated technology and will also take into account possible redundancy to deal with radiation effects (e.g. SEU).  

\begin{figure}[htbp]
  \centering
  \includegraphics [width=.65\textwidth]{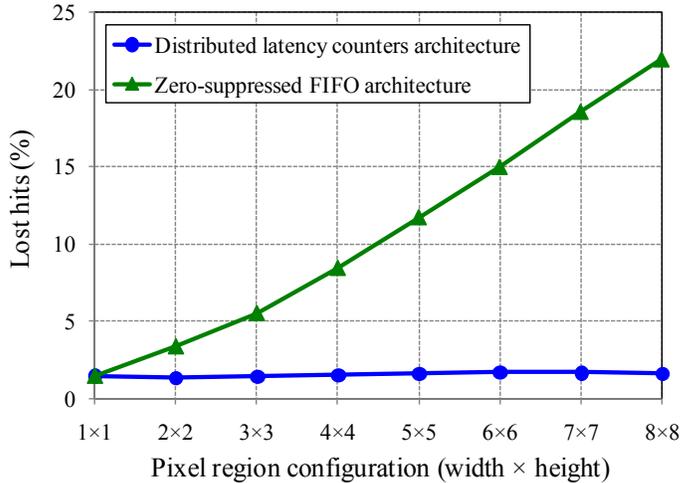}
  \caption{Pixel hit loss as a function of PR size for the zero-suppressed FIFO (in green) and distributed latency counters (in blue) buffering architectures.}
  \label{fig:losthits}
\end{figure}

%
Results related to hit loss are plotted in Figure~\ref{fig:losthits}. There is a remarkable difference between the two architectures as the PR size increases. This is related to the fact that in the zero-suppressed FIFO architecture the regional logic keeps the whole PR busy while the hit pixels are measuring the hit ToT, causing a dead time during which the idle PUCs of the region are prevented from processing hits coming from following bunch crossings. For this reason bigger PRs which get more input hits are more affected by such dead time. The distributed latency counters architecture, on the other hand, is such that the PUCs of the region process the hits of a cluster independently from each other and there is no dead time related to the fact that the whole PR is busy. Therefore, at behavioral level, this is the architecture to be preferred. As a final remark it should be underlined that because buffers have been oversized for collection of statistics on occupancy, no hits in these results are lost because of buffer overflow. 

\section{Conclusions}
\label{sec:conclusions}
A simulation and verification framework has been implemented as a development tool for the next generation pixel chips of the ATLAS and CMS experiments at the LHC. The environment, coded using the SystemVerilog hardware description and verification language and the UVM class library, supports both high level system simulation and IP simulation. The verification components of the environment are in charge of generating the input hits for the pixel chip DUT, monitoring its input and output signals and performing conformity checks between input and output and statistics collection related to its performance. The hit generator section, in particular, can produce different classes of parameterized hits in order to obtain different cluster shapes on the pixel chip matrix.

The environment has been tested for a comparative study of buffering architectures in pixel regions. A fully shared zero-suppressed FIFO architecture and a distributed one based on shared latency counters have been described at behavioral level and simulated in order to collect statistics on memory occupancy and hit loss. While both the architectures have shown a comparable behavior from the number of buffering locations point of view, the distributed one has featured a hit loss rate independent from the number of pixels grouped in a region.

Further developments on the environment are related to the hit generation section. In particular, it should be introduced the possibility of integrating hit patterns from physics simulations to the current generation scheme; moreover, additional distributions which are not standard for SystemVerilog are being developed as constraints for the charge of the hits to be generated (e.g. Landau distribution). The hit generator will also be refined with the randomization of the delays, enabling the simulation of jitters at different clock frequencies. Direct interfacing of the enviornment with the ROOT framework, for both the hit generation and the analysis, is also being considered. 
This development will also enable further architectural studies of pixel chip building blocks, e.g. column bus arbitration schemes; it is in fact a long-term goal to enhance the existent environment to use it as a design tool for low power and fault tolerance architecture optimization.

\acknowledgments

The authors would like to thank Dr. Tuomas Poikela for the technical support and the fruitful discussions.

\end{document}